\documentclass[aps,prl,twocolumn,showpacs]{revtex4}
\usepackage{graphicx}
\begin{document}
\title{Geometric phases and Wannier functions of Bloch electrons in 1-dimension}
\author{Joydeep Bhattacharjee and Umesh V Waghmare}
\affiliation{Theoretical Sciences Unit,
Jawaharlal Nehru Centre for Advanced Scientific Research,\\ Jakkur PO,
Bangalore 560 064, India}
%==========================================================================================
\begin{abstract}
We present a formal expression for Wannier functions of composite bands of 1-D 
Bloch electrons in terms of parallel-transported Bloch functions and their 
non-Abelian geometric phases. Spatial decay properties of these Wannier functions 
are studied in the case of simple bands of 1-D model insulator and metal. Within 
first-principles density functional theory, we illustrate the formalism through the
construction of Wannier functions of polyethylene and polyacetylene. 
\end{abstract}
\maketitle
%===================================================================================
%\section{Introduction}

Since the introduction in 1937, Wannier functions (WF)\cite{Wan} have been 
conceptually interesting and useful in studies of electronic properties of materials. 
They highlight the atomic orbital character of electrons and chemical bonding\cite{kohn73}
and appear naturally in the symmetry and analytical properties of energy bands 
in solids\cite{descloizeux, Zak89}. They form an excellent basis set for use 
in the construction of first-principles model Hamiltonians in studies of
phase transitions\cite{rabe95,ole}. There is growing interest in WFs as they
provide a localized basis in real space for the description of electronic states 
and are useful in the implementation of computationally efficient``linear scaling"
algorithms \cite{galli, goedecker}. 

WFs are obtained as a unitary transformation of Bloch functions\cite{Wan}.
A relationship between the first moment of squared WFs (density) and geometric phases\cite{berry} of 
Bloch electron was uncovered in the context of adiabatic dynamics of Bloch electron\cite{Zak} and
later in the modern theory of polarization\cite{KSV, Resta}. WFs are {\it not}
unique due to the freedom in choice of phases accompanying Bloch functions, but they are
desired to be highly localized for practical applications. The second moment (cumulant) of
squared WFs, a measure of their spatial range, is used as an objective function for 
minimization in the variational procedure to construct WFs developed by 
Marzari and Vanderbilt\cite{MV}.

In this paper, we focus on the solution of WF problem for systems periodic in 1-D,
and comment on its relation and extension to higher dimensions.
We use Bloch functions parallel transported along the Bloch vector {\bf k}
in the Brillouin zone as a starting point in the construction of WFs.
As these are not periodic in reciprocal space, we present an explicit form for a unitary 
transformation to construct a set that is periodic as a function of {\bf k}. For the 1-D case,
we show that the WFs obtained as a Fourier transform of this set are
maximally localized.

Starting with a subspace $S_{k_0}$ of $N$ energy eigenfunctions at $k_0$, 
we use parallel transport to obtain Bloch
functions $e^{ikx} u^{||}_{kn}(x)$ for $k \in ( k_0, k_0+ \frac{2 \pi}{a})$, $a$ 
being the lattice constant and $n$ being an index for electronic states. 
This is accomplished using $|\frac{\partial u_{kn}}{\partial k}>$ obtained routinely
within the parallel transport gauge:
\begin{equation}
\langle u^{||}_{km}|\frac{\partial u^{||}_{kn}}{\partial k} \rangle=0
\label{paralleltrnpt}
\end{equation}
in density functional theory linear 
response\cite{BGT, GAT} (or $k.p$ perturbation theory) calculations.
We use second-order Runge-Kutta procedure\cite{recipes} to translate from 
$u^{||}_{kn}(x)$ to $u^{||}_{k+\Delta k n}(x)$. Geometric phases, generalized for
degenerate states\cite{WilZee} and paths closed within a gauge transformation\cite{Zak},
form an $N\times N$ Hermitian matrix $\Gamma$ and are given by:
\begin{equation}
\left[e^{i \Gamma}\right]_{mn} = \langle u_{m k_0}^{||} 
\vert \exp\left(i\frac{2 \pi x}{a}\right)
\vert u_{n k_0+\frac{2\pi}{a}}^{||} \rangle 
\label{nonabelianGammaPar}
\end{equation}
The electronic part of the macroscopic polarization\cite{KSV} $P$ is readily given 
by $\frac{2\pi e}{a}Tr(\Gamma)$.
\iffalse by $\frac{e}{\Omega}Tr(\Gamma)$.
 $\Omega$ being the unit cell volume \fi

Non-diagonal form of the matrix $\Gamma$, in general, indicates that a starting 
wavefunction at $k_0$ mixes with other members of the subspace $S_{k_0}$ as $k$ is 
transported to $k_0+\frac{2\pi}{a}$, amounting to a unitary transformation of the 
subspace $S_{k_0}$. Diagonalization of $\Gamma$ and corresponding transformation 
of Bloch functions
in $S_{k_0}$ result in states that are decoupled from each other during parallel transport, 
effectively reducing the problem of composite bands to $N$ simple bands. If $M$ forms
an $N\times N$ matrix whose columns are the eigenvectors of $\Gamma$, we obtain
an auxiliary set of parallel transported wavefunctions $\tilde{u}_{nk}$:
\begin{equation}
\vert \tilde{u}_{nk} \rangle = \sum^N_m M^{\star}_{mn} \vert u^{||}_{mk} \rangle
\label{diagonalspace}
\end{equation}
where $ \gamma_n \delta_{m n} = (M^{\dagger} \Gamma M)_{m n} $. The resulting wave functions 
$\left\{|\tilde{u}_{nk}\rangle\right\}$ are independent of the choice of $k_0$ 
and satisfy:
$\langle \tilde{u}_{mk_0}|\exp{\left(i\frac{2\pi x}{a}\right)}|\tilde{u}_{n(k_0+2\pi/a)}\rangle
=\exp{\left(i\gamma_n\right)}\delta_{mn}$ where $\gamma_n$ is the $n$-th eigen value 
of $\Gamma$ and $\gamma_n\frac{a}{2\pi}=<x>_n$.
%================FIGURE====================================================================
\begin{figure}[b]
\centering
\includegraphics[scale=0.45]{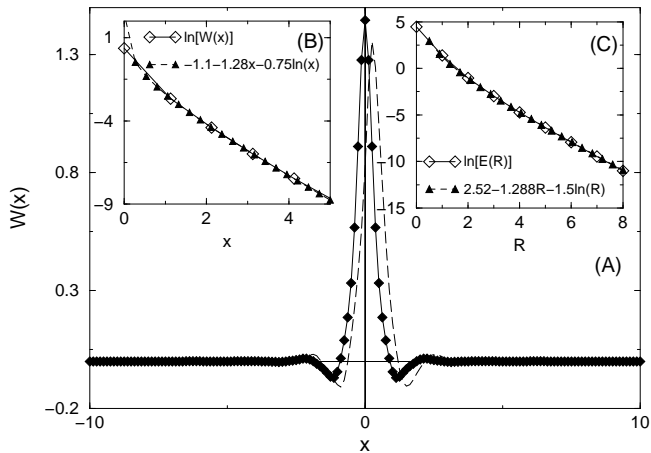}
\caption{(A) WF constructed using (\ref{finaleqni}), for the lowest 
non-degenerate band (n=1) in case of 1D insulator  modeled with: 
Gaussian centro-symmetric potential (solid line)
with $ c=-10, b=0.3$ and asymmetric potential (dashed line) with 
$ d_1=-0.3, d_2=-0.2, c_1=-5.0, -c_2=-3.0$, both  with $ a=1.0 $.
(B) and (C) shows decay of WF and E(R) respectively for the Gaussian centro-symmetric potential. 
The power law exponents appear as the coefficients of ln(x)  and  ln(R) in (B) and (C) respectively
(all in atomic units).}
\label{HeVanderbilt}
\end{figure}
%==========================================================================================
%=========TABLE================================== T A B L E ================================
\begin{table}[t]
\caption{Polarization $<x>$ and squared localization length $\ell^2$ calculated for a 
single isolated band $(n=1)$ for different potentials: 
(1) Gaussian potential with $c=-10.0, d=0.3 $; 
 and (2) asymmetric potential with $d_1=-0.3, d_2=-0.2, c_1=-5.0, -c_2=-3.0$ both with $a=1.0$
(all in atomic units).}
%\begin{center}
\begin{tabular*}{3.4in}{c@{\extracolsep{\fill}}c@{\extracolsep{\fill}}c@{\extracolsep{\fill}}
c@{\extracolsep{\fill}}c@{\extracolsep{\fill}}}
 \hline 
\hline
V(x) & $<x>$  & $\ell^2_{WF}$ & $\ell^2_{LR}$ & $\ell^2_{BF}$ \\
\hline
(1) & $ 0.0 $   & $ 0.305 $   & $ 0.305 $ & $ 0.305 $  \\
%\hline
(2) & $ 0.288 $ & $ 0.484 $ & $ 0.484 $ & $ 0.484 $  \\  
\hline
\hline
\end{tabular*} 
%\end{center}
\label{LocLengthCompare}\\
\end{table}
%==========================================================================================
%================FIGURE====================================================================
\begin{figure*}[t]
\centering
\includegraphics[scale=0.26]{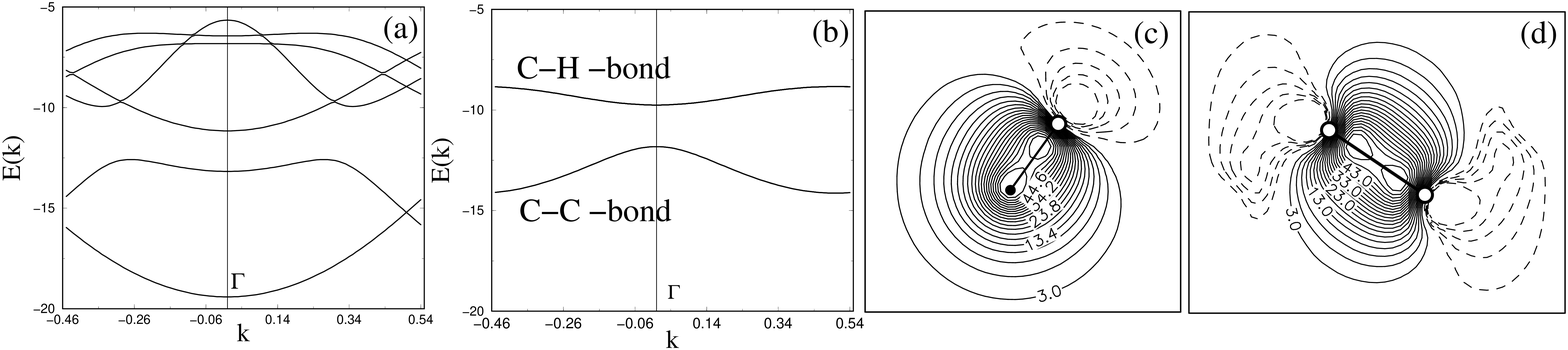}
\caption{Polyethylene: (a) Band structure, 
(b) Transformed band structure of $\langle \tilde{u}_{nk}|H(k)| \tilde{u}_{nk}\rangle$,
(c) WF corresponding to C-H bond,
(d) WF corresponding to C-C bond. }
\label{polyethylene}
\end{figure*}
%==========================================================================================
%================FIGURE====================================================================
\begin{figure*}[t]
\centering
\includegraphics[scale=0.28]{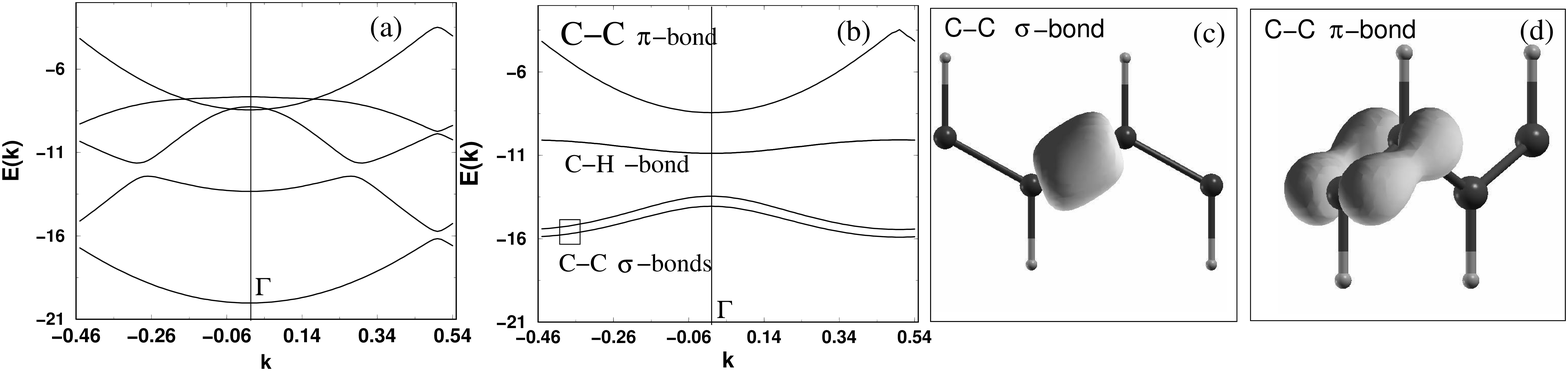}
\caption{Polyacetylene: (a) Band structure, 
(b) Transformed band structure of $\langle \tilde{u}_{nk}|H(k)| \tilde{u}_{nk}\rangle$,
(c) WF corresponding to C-C $ \sigma-$bond,
(d) WF corresponding to C-C $ \pi-$bond. In (c) and (d) the isosurfaces are prepared
using XCrysDen software\cite{xcrysden}.}
\label{polyacetylene}
\end{figure*}
%==========================================================================================
Wavefunctions $\left\{|\tilde{u}_{nk}\rangle\right\}$ are not periodic in $k$ with a period 
$\frac{2\pi}{a}$ if $\gamma_n\not=0$.
A simple U(1) transformation can be used to derive $\left\{|v_{nk}\rangle\right\}$
that have the lattice periodicity in both reciprocal and real space:
\begin{equation}
|v_{nk}\rangle = \exp \left\{-ik\left(\frac{\gamma_n a}{2\pi}\right)\right\}|\tilde{u}_{nk}\rangle
\label{Solution2}
\end{equation} 

%================FIGURE====================================================================
\begin{figure}[b]
\centering
\includegraphics[scale=0.67]{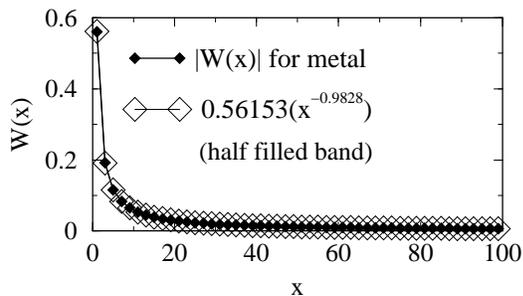}
\caption{Decay of WF for metal with half filled band 
modeled by a centro-symmetric potential at T=0K:
$ d1=0.25, c1=-5.0, d2=0, c2=0 $ with $ a=1.0 $ ( all in atomic units ).}
\label{metalWanDecay}
\end{figure}
%==========================================================================================
In terms of parallel transported wave functions,
$|v_{nk}\rangle = \sum_m U_{nm,k}|u^{||}_{mk}\rangle$,
where
\begin{equation}
U_{nm,k} =\exp \left\{-ik\left(\frac{\gamma_n a}{2\pi}\right)\right\} M^{\star}_{mn} 
\label{nettransformation}
\end{equation}

In the frame-work of differential geometry\cite{SBh} the transformation $U_{nm,k}$  
projects the \{$|u^{||}_{mk}\rangle$ \} in the ray space onto \{$|v_{nk}\rangle$ \}
in the physical space. Each function $|v_{nk}\rangle$ is  periodic and continuous in
$k-$space and can be Fourier transformed to obtain the WF $W_n(x, R)$: 
\begin{eqnarray}
|W_n(x,R)\rangle&=&\sum_{k} e^{ik(x-R)} |v_{nk}\rangle \nonumber \\
&=&\sum_{k,r} e^{ik(x-R)} C_{nr} |u_{rk}\rangle
\label{finaleqni}
\end{eqnarray}
where $\left\{|u_{rk}\rangle\right\}$ are the energy eigenfunctions obtained from
diagonalization and $C_{nr}=\langle v_{nk}|u_{rk}\rangle$.
Eqn (\ref{finaleqni}) is the main formal result of this paper that gives WF
in the general case of composite bands for systems that are periodic in one dimension.
Using Eqn (\ref{paralleltrnpt}) and (\ref{nettransformation}), these WFs 
can be shown to have two important properties (in addition to being orthonormal): 
\begin{eqnarray}
1.& &\int x W^{\star}_n(x,0)W_n(x,0) dx = \gamma_n \nonumber\\
2.& &\ell^2_{WF,n} = \int x^2 W^{\star}_n(x,0)W_n(x,0) dx - \gamma_n^2
\label{omega}
\end{eqnarray}
squared localization length $\ell^2_{WF,n}$ is independent 
of phase factors accompanying the energy eigenfunctions and can be shown to be
equal to the one obtained using linear response\cite{viethen, Waghmare}. 
Thus, these WFs are the same as maximally localized Wannier 
functions\cite{MV} in 1-dimension.
A general form of $U(1)$ transformation in Eqn (\ref{Solution2}) is 
$\exp \left\{-ik\left(\frac{\gamma_n a}{2\pi}\right)-i k.La-i g(k)\right\}$, 
where $L$ is an integer and $g(k)$ a periodic function of k.
It is readily shown that the term $k.La$ in the exponent alters\cite{KSV} $<x>_n$ 
by $La$ but keeps $\ell^2_{WF,n}$ unchanged. 
On the other hand, the term $g(k)$ keeps $<x>_n$ unchanged, but increases $\ell^2_{WF,n}$.

We propose a generalization of expression (\ref{finaleqni}) to metals by 
including the Fermi-Dirac distribution factor $f(E_{nk},T)$:
\begin{equation}
|W^T_n(x,R)\rangle=\sum_{k,r} e^{ik(x-R)} C_{nr}\left[f(E_{rk},T)\right]^{\frac{1}{2}}
|u_{rk}\rangle
\label{finaleqn}
\end{equation}
These generalized WFs do not form an orthonormal set in general, but yield the same form
of density matrix for metals as that for insulators:
$$\rho^T(x,x')=\sum_{nR}W^{T \star}_n(x,R)W^T_n(x',R).$$
Secondly, they can be shown to have infinite localization length in the case of partially
filled bands, arising from the discontinuity in  $f(E_{nk},T)$ at the Fermi energy\cite{BeigiArias}.

We first use above formalism to construct WFs for an isolated single band 
in 1D model insulator and metal (partially filled band). We consider two
generic types of model potentials:
(1) Gaussian centro-symmetric potential: 
$ V(x)=\sum_{n=1}^{\infty}\frac{c}{d\sqrt{2\pi}}exp[-\left(x-na\right)^2/d^2] $, which was used
by He and Vanderbilt\cite{HeVanderbilt} in analysis of spatial decay of WFs 
and is used to benchmark our calculation, and (2) general asymmetric potential:
$V(x)=c_1\left[1+\cos\left\{2\pi(x+d_1)/a\right\}\right]+c_2\left[1+\cos\left\{4\pi(x+d_2)/a
\right\}\right]$, which we use to test the universality\cite{HeVanderbilt} of decay exponents 
of WFs. Fig.\ref{HeVanderbilt}(A) shows WFs constructed using our formalism for the two
potentials.

We calculate the squared localization length $\ell^2_{n}$ using three different methods:
$\ell^2_{WF,n}$ using Eqn (\ref{omega}), $\ell^2_{LR,n}= \langle \frac{\partial u_{kn}}
{\partial k} \vert \frac{\partial u_{kn}}{\partial k} \rangle$ obtained using linear
response and
$\ell^2_{BF,n} =\frac{2}{\triangle k^2}\sum_{BZ} (1-Re\langle u^{||}_{n,k} | u^{||}
_{n,k+\triangle k}\rangle) \triangle k $, calculated from the parallel transported 
Bloch functions.
As summarized in Table \ref{LocLengthCompare}, $\ell^2$ calculated using the three methods
for both the potentials studied here are identical within numerical errors.
As the $\ell^2_{LR,n}$ is known to be gauge invariant for a given subspace\cite{viethen},
and correspond to maximum localization, this numerically verifies maximal localization of the
WFs constructed using the present formalism for the two cases.

We now compare the spatial decay behavior of the present WFs with the maximally
localized WFs\cite{MV} which exhibit a power law decay\cite{HeVanderbilt}
in addition to the exponential one\cite{Kohn}: $ W_n(x)\approx x^{-3/4}e^{hx} $,
where $ h $ is defined from the $k-$ point $ \pi/a+i h $ at which $E_0=E_1$ in the complex band structure.
Similarly $ E_n(R) $, defined as $ E_n(R)=\langle W_n(x,R) |{H}| W_n(x,0)\rangle $, decays as:
$ E(R)\approx x^{-3/2}e^{hR}$.
The power law decay exponents $ -3/4 $ for $ W_n(x,R) $ and $ -3/2 $ for $ E_n(R) $ are universal 
numbers\cite{HeVanderbilt}. Calculating $ h $ from the complex band structure and
fitting these general decay forms to the WFs in this work, we find the same power law 
exponents for both the model potentials considered. 
Results for the centro-symmetric potential are shown in Fig.\ref{HeVanderbilt}(B) and (C).

WF constructed using Eqn. (\ref{finaleqn}) for a partially filled band
($T=0K$) is shown in Fig.\ref{metalWanDecay}. Form of the spatial decay of this WF 
agrees with that of the density matrix ($\propto x^{-\eta}$) predicted analytically 
by Beigi and Arias\cite{BeigiArias}. We observe that the 
exponent $\eta$ is between 0.95 and 1 for any fraction of band filling at 0K. This means
the localization length of metals being infinity.

We now illustrate the present method within first-principles density functional theory 
in the general case of composite bands by constructing WFs for 
two polymer chain molecules: polyethylene (PE) and polyacetylene (PA). 
A unit cell of a polyethylene (polyacetylene) chain consists of 2[CH$_2$] (2[CH]) with twelve 
(ten) valence electrons occupying the lowest six (five) bands. The unit cell considered is large 
enough in the transverse directions to minimize the interaction between the periodic images of the chain.
We use standard density functional theory code\cite{castep} to carry out these calculations. We used
100 Runge-Kutta steps in the parallel transport and 20 $k-$points in the Fourier transform (Eqn (\ref{finaleqni})) to obtain WFs.

The electronic structure of PE [Fig.\ref{polyethylene}(a)] apparently groups into two non-crossing sets of 
bands (the lowest two and the rest). Interestingly, unitary matrix $ M $ we determined from the 
geometric phases mixes bands from these groups. We use $M$ to transform the $u_{nk}$ to $\tilde{u}_{nk}$, 
and plot a transformed band structure [Fig.\ref{polyethylene}(b)] with expectation values of energy 
of $\tilde{u}_{nk}$.
We find two distinct bands: (1) the lower one with double degeneracy corresponding to C-C bonds and
 (2) the upper one with quadruple degeneracy corresponding to C-H bonds.
In Table \ref{polyethdata} we summarize $<\vec{r}>$ and $\ell^2$ values for the six WFs
in two groups corresponding to the two transformed bands (1) and (2).
\iffalse
We find two distinct bands, lower one with double degeneracy corresponding to C-C bonds and
upper one with quadruple degeneracy corresponding to C-H bonds.
 We summarize 
$<\vec{r}>$ and $\ell^2$ values for the six WFs in Table \ref{polyethdata}.
\fi
The average squared localization length $ \bar{\ell}^2_{WF}=(1/N)\sum^N_n \ell^2_{WF,n} $ is 0.23 $\AA^2$ and
agrees well with  linear response calculation $\bar{\ell}^2_{LR}$ of value 0.23 $\AA^2$ indicating   
maximal localization. Contour-plots of the two WFs are shown in
Fig.\ref{polyethylene}(c,d)
%=========TABLE================================== T A B L E ================================
\begin{table}[t]
\caption{Polarization $<\vec{r}>$ and squared localization length $\ell^2_{WF}$ of polyethylene 
WFs (all lengths are in \AA).}
\begin{tabular}{ccccccccccc} \hline
\hline
\multicolumn{1}{c}{Bond} & \multicolumn{9}{c}{$<\vec{r}>$} 
& \multicolumn{1} {c} {$\ell^2_{WF}$} \\ \hline
C-C& &(0.0,0.0,0.0)& & &(1.25.0,0.0)& & & & &0.24 \\
C-H& &($\mp$0.63,0.53,$\pm$0.91)& & &($\pm$0.63,-0.53,$\mp$0.73)& & & & &0.23 \\ 
\hline 
\hline
%\multicolumn{11}{c}{All lengths are in  \AA  }
\end{tabular} 
%\end{center}
\label{polyethdata}
\end{table}
%==========================================================================================

Similarly, electronic structure of PA (Fig.\ref{polyacetylene}(a)) yields 
the transformed band structure (Fig.\ref{polyacetylene}(b)) consisting of 
four isolated bands: two of them (lowest in energy) are almost degenerate corresponding to the two C-C 
$\sigma$-bonds of slightly different lengths. The one in the middle is doubly degenerate and corresponds to
the two C-H bonds and the one highest in energy corresponds to the C-C $ \pi$-bond.
Fig.\ref{polyacetylene}(c,d) shows isosurfaces of WFs describing C-C $\sigma$ bond 
($\ell^2_{WF}=0.22\AA^2$) and C-C $\pi$ bond ($\ell^2_{WF}=2.95\AA^2$) in polyacetylene.
An order of magnitude higher $\ell^2_{WF}$ of C-C $\pi$ bonds than the others 
makes $\pi$-conjugated systems particularly interesting for transport properties.

While the logic behind the present formalism is the construction of simple continuous and periodic
functions from parallel transported Bloch functions, the resulting WFs in 1-D are maximally
localized. Its generalization to higher dimensions is natural when the geometric phases $\Gamma_{\alpha}$
for paths along different directions in the Brillouin zone commute. In most cases, however,
the  $\Gamma_{\alpha}$'s do not commute (corresponding to nontrivial phases for closed paths within the
Brillouin zone) implying the WFs can not be perfectly localized in all the
three directions simultaneously.

In conclusion, we have presented a formal expression for WFs of 1-d Bloch
electrons and a first-principles method for their construction through calculation of geometric phase matrices $\Gamma$.
It can be readily implemented in any DFT code that includes linear response. It naturally identifies 
the groups of bands that make up WFs (or bonds) through the eigenvectors of the
$\Gamma$ and their polarization through eigenvalues of $\Gamma$. The method should be very useful in
studies of 1-D periodic systems in nano-science.

\acknowledgements
UVW thanks Karin M. Rabe for many useful discussions.
We also thank G. Baskaran, J. Samuel and Ayan Datta for stimulating discussions. 
JB thanks CSIR, India for a research fellowship.
 This work was supported by the J Nehru Centre for Advanced Scientific Research, funded
by the Department of Science and Technology, Government of India.
%==================== REFERENCES =========================================================

\end{document}